\documentclass{acm_proc_article-sp}
\usepackage{graphicx}
\usepackage{algorithmic}
\usepackage{algorithm}
\usepackage{ amsmath}

\long\def\remove#1{}
\newcommand{\myparagraph}[1]{ {\bfseries{\em #1:}} \normalsize}
\newcommand{\denselist}{
     \setlength{\itemsep}{0pt}
     \setlength{\parsep}{1.5pt}
     \setlength{\topsep}{1.5pt}
     \setlength{\parskip}{2pt}
     \setlength{\partopsep}{0pt}
     \setlength{\labelwidth}{1em}
     \setlength{\labelsep}{0.5em} }

\newcommand{\bdesc}{\begin{description}\denselist}
\newcommand{\edesc}{\end{description}}

\newcommand{\blist}{\begin{itemize}\denselist}
\newcommand{\elist}{\end{itemize}}
\begin{document}
\conferenceinfo{MLG}{'10 Washington, DC USA}
\CopyrightYear{2010}
\crdata{978-1-4503-0214-2}

\title{Centrality Metric for Dynamic Networks
%
%
}

\numberofauthors{3}

\author{
\alignauthor
Kristina Lerman\\
       \affaddr{USC Information Sciences Institute}\\
       \affaddr{4676 Admiralty Way}\\
       \affaddr{Marina del Rey, CA 90292}\\
       \email{lerman@isi.edu}
\alignauthor Rumi Ghosh\\
       \affaddr{USC Information Sciences Institute}\\
       \affaddr{4676 Admiralty Way}\\
       \affaddr{Marina del Rey, CA 90292}\\
       \email{rumig@usc.edu}
\alignauthor
Jeon Hyung Kang\\ 
       \affaddr{USC Information Sciences Institute}\\
       \affaddr{4676 Admiralty Way}\\
       \affaddr{Marina del Rey, CA 90292}\\
       \email{jeonhyuk@usc.edu}
}
\date{}

\maketitle
\begin{abstract}
Centrality is an important notion in network analysis and is used to measure the degree to which network structure contributes to the importance of a node in a network. While many different centrality measures exist, most of them apply to static networks.
Most networks, on the other hand, are dynamic in nature, evolving over time through the addition or deletion of nodes and edges.
A popular approach to analyzing such networks represents them by a static network that aggregates all edges observed over some time period. This approach, however, under or overestimates centrality of some nodes. We address this problem by introducing a novel centrality metric for dynamic network analysis. This metric exploits an intuition that in order for one node in a dynamic network to influence another over some period of time, there must exist a path that connects the source and destination nodes through intermediaries at different times. We demonstrate on an example network that the proposed metric leads to a very different ranking than analysis of an equivalent static network.
We use dynamic centrality to study a dynamic citations network and contrast results to those reached by static network analysis.
\end{abstract}



\keywords{dynamic networks, centrality, networks} 

\section{Introduction}
The structure of many complex systems, from biological and social systems to the World Wide Web and more recently the Social Web, can be represented as a network. Ability to analyze networks in order to identify important nodes and discover hidden structure has led to important scientific and technological breakthroughs. As a single profound example, PageRank~\cite{PageRank} algorithm, which ranks Web documents by analyzing the structure of hyperlinks between them, has revolutionized both Internet search and commerce. Network analysis algorithms are also used to discover communities of like-minded individuals~\cite{Newman04}, identify influential people~\cite{Kempe03} and blogs~\cite{Leskovec07kdd}, rank scientists~\cite{Radicchi09} and find important scientific papers~\cite{Walker06,Chen07,Sayyadi09sdm}. With few exceptions, these metrics and algorithms have been applied to \emph{static} networks. Real-world networks, however, are \emph{dynamic} in nature, because their topology can change over time with addition of new nodes and edges or removal of existing ones.



This paper defines a novel centrality metric for dynamic networks. The metric generalizes the path-based centrality used in network analysis~\cite{Bonacich:2001,Ghosh09socialcom} which measures centrality of a node by the number of paths, of any length, that connect it to other nodes. The dynamic centrality metric exploits an intuition that in order for a message sent by one node in a network to reach another after some period of time, there must exist a path that connects the source and destination nodes through intermediaries at different times. A distinctive feature of this metric is that it is parameterized by factors that set both time and length scale of interactions. These parameters can in some cases be estimated from data.
We use dynamic centrality to rank nodes by the number of time-dependent paths that connect them to other nodes in the network. In addition to discovering best connected, or influential, nodes, the method can identify nodes that are most connected to a specific node and, therefore, have highest influence on it. We perform detailed analysis of a toy dynamic network and show that dynamic network analysis can lead to a vastly different ranking than analysis of an equivalent static network. We also study a real-world dynamic network that represents scientific citations data set. We find optimal parameters for the metric by fitting it to the citation chains' temporal and length distributions. We show that dynamic centrality can produce a radically different view of what the important nodes in the network are than static measures and leads to new insights about the structure of the dynamic network.

In Section~\ref{sec:related} we review existing research on network analysis and identify challenges in extending it to dynamic networks. We define dynamic centrality in Section~\ref{sec:dynamic} and present mathematical formalism that allows us to compute it from the snapshots of the network over time. We demonstrate in Section~\ref{sec:ranking} how this metric can be used to rank nodes in a dynamic network. In Sec.~\ref{sec:citations} we apply dynamic centrality to study the scientific papers citations network and show that dynamic centrality can lead to a drastically different view of importance than analysis performed on an equivalent static network.

\section{Background and Related Work}
\label{sec:related}
\myparagraph{Centrality metrics}
Centrality determines node's importance in a network. This measure is dependent on the network structure.
The simplest centrality metric, degree centrality, measures the number of edges that connect a node to other nodes in a network. Over the years many more complex centrality metrics have been proposed and studied, including Katz status score~\cite{Katz}, $\alpha$-centrality\cite{Bonacich:2001}, betweenness centrality~\cite{Freeman79}, and several variants based on random walk~\cite{Stephenson89,Noh02,Newman05}, the most famous of which is PageRank~\cite{PageRank}. The path-based centrality metrics \cite{Bonacich:2001} measure the extent to which a node can influence, or control how much information flows to, other nodes in a network.

Consider, specifically, $\alpha$-centrality defined by Bonacich\cite{Bonacich:2001}, which measures the total number of attenuated paths of any length between nodes $i$ and $j$. Let $A$ be the adjacency matrix of a network, such that $A_{ij}=1$ if an edge exists from $i$ to $j$ and $A_{ij}=0$ otherwise. \emph{$\alpha$-centrality matrix} is given by:
\begin{equation}
\label{eq:static}
C^s(\alpha,\beta) = \beta A +\beta \alpha A \cdot A + \cdots + \beta \alpha^n  A^{n+1} \cdots
\end{equation}
\noindent
where $\beta$ is the attenuation factor along a direct edge (from the originating node) in a path, and $\alpha$ is the attenuation factor along an indirect edge (from any intermediate node) in a path. Although attenuation factors along subsequent edges in a path could in principle be different, for simplicity, we take them all to be the same, namely $\alpha$. The first term in the equation above gives the number of paths of length one (edges) from nodes $i$ to $j$, the second the number of paths of length two, and so on.

The tunable parameter $\alpha$ sets the length scale of interactions.
For $\alpha=0$, $\alpha$-centrality takes into account direct edges only and reduces to degree centrality (weighted by $\beta$). As $\alpha$ increases, $C^s(\alpha,\beta)$ becomes a more global measure, taking into account more distant interactions. Nodes can be ranked according to the number of paths that connect them to other nodes. In previous work~\cite{Ghosh08snakdd,Ghosh09socialcom} we used this framework to identify both locally and globally influential nodes, as well as discover community structure of networks.

\myparagraph{Dynamic networks}
While most of network analysis research focused on \emph{static} networks, recently researchers began to study \emph{dynamic} networks, whose topology changes in time through the addition or removal of nodes and edges.
\cite{Braha06} represented a dynamic network by time series, or snapshots, of the network, each of which aggregates links over a time scale much shorter than the entire observation period. They studied how degree centrality evolves in a dynamic network.
\cite{Eckmann04} observed that activation of links in a dynamic network creates a flow of information that leads to coherent clusters. They introduced a metric to study these structures and their evolution. The metric modifies the traditional clustering coefficient. Specifically, it measures the number of triangles in which a node of degree $v$ participates.
Similarly, \cite{BergerWolf06} proposed a formal framework for identifying communities within dynamic networks based on the temporal structure of underlying interactions.
Our focus in this paper is not to identify coherent structures or groups in a dynamic network. Instead, we want to define an intuitive metric that enables us to rank nodes in a network. We generalize $\alpha$-centrality to dynamic networks. Using this metric we can  rank nodes by how well they are connected to other nodes in the network \emph{through time}, thereby identifying important or influential nodes.

\myparagraph{Time-aware ranking}
Closely related to dynamic network analysis is the problem of time-aware ranking of Web pages in information retrieval. This research is motivated by the observation~\cite{Baeza-Yates02} that PageRank's Web ranking algorithm  is biased against newer pages, which may not have had enough time to accumulate links to give it a high rank. Several methods have been proposed to address the recency bias in PageRank, including \cite{Baeza-Yates02,Yu04,Berberich05,Dong10wsdm}. In general terms, these methods weigh edges in the network by age, with newer edges contributing more heavily to a page's importance. Our motivation is different. Rather than focus on improving the rank of newer nodes, we focus instead on defining a time-aware centrality metric that takes the temporal order of edges into account.

Authors of \cite{OMadadhain05} considered the temporal order of edges in the flow of information on a network. They proposed EventRank algorithm, a modification of PageRank, that takes into account a temporal sequence of events, e.g., spread of an email message, in order to calculate importance of nodes in a network. This approach takes into account the effect of the \emph{dynamic process} on ranking. In contrast, we consider the effect of the \emph{dynamic network topology} on ranking. These approaches are somewhat related: our method can be said to estimate the expected value of all temporal sequences taking place on the network.

\myparagraph{Scientific citations}
Ranking scientific publications is an interesting application for dynamic network analysis. A long line of bibliometrics research attempted to define objective metrics for identifying important scientific papers, researchers, publication venues, and institutions. The now-accepted measures for evaluating the impact of papers and individual researchers include citations count and h-index~\cite{hindex}. A breakthrough in this field came with the representation of the body of scientific literature as a multi-partite network consisting of authors, papers, and publication venues, where a link between an author and a paper denotes a researcher's authorship of the paper, a link between two papers indicates a scientific citation, etc. This representation allows the structure of the network to be considered in ranking papers and authors.

Scientific papers citations data set can also be considered a dynamic network, in which newly published papers create edges to existing papers by citing them. Unlike a generic dynamic network, however, edges in a citations network are never destroyed.
All previous work treated a citations network essentially as a static network that {aggregates} all citations links created over some time period.
\cite{Chen07} implemented PageRank algorithm on such an aggregated  network to find most influential papers. \cite{Radicchi09} divided the entire data period into homogeneous intervals containing equal numbers of citations and applied a PageRank-like algorithm to rank papers and authors within each time slice, thereby, enabling them to study how an author's influence changes in time. In order to address PageRank's bias for older papers, \cite{Walker06} introduced {CiteRank}, a modified version of PageRank, that explicitly takes paper's age into account.  {CiteRank}  performs a random walk on a citations graph, but initiates the walk from a recent paper $i$ chosen randomly with probability $p_i=e^{\frac{age_i}{\tau}}$, where $age_i$ is the age of the paper and $\tau$ characteristic decay time. The random walk, however, was performed on an aggregated network. Authors estimated parameters of the random walk by fitting papers'  {CiteRank} score to the number of citations accrued by them over some time period. \cite{Sayyadi09sdm} described \emph{FutureRank}, an algorithm that predicts paper's PageRank scores some time in the future.  {FutureRank} implicitly takes time into account by partitioning data in time, and using data in one period to predict paper's ranking in the next. Similar to \cite{Radicchi09}'s approach,  {FutureRank} combines influence rankings computed on the papers and authors networks into a single score. This score is shown to correlate well with the paper's PageRank score computed on citations links that will appear in \emph{the future}.
However, no previous method took the temporal order of citations edges into account. The method proposed in this paper, on the other hand, ranks scientific publications by explicitly taking temporal constraints on citations links into account.


\section{Dynamic Centrality Metric}
\label{sec:dynamic}
A \textit{dynamic} network as a network whose topology changes over time through addition or removal of edges.
Let $t$ be the smallest time interval in which there is no change in the topology of the network. Following \cite{Braha06}, we represent network at time $t_{i} (i \in 1,\cdots, n)$ by a graph $G_{t_{i}}= (V_{t_{i}},E_{t_{i}} )$ with $V_{t_{i}}$ nodes and $E_{t_{i}}$ edges between them at time $t_{i}$. We define $ A(t_{i})$ as the adjacency matrix corresponding to $G_{t_{i}}$. A  time series of network snapshots $G_{t_{1}}, G_{t_{2}}, \cdots G_{t_{n}} $ (where $t_{i}-t_{i-1} \le t $) could then be used to represent a dynamic network over the time period $\Delta_{1,n}=\{t_{1} \ldots t_{n}\}$.

\remove{
We assume that a given dynamic network has the following properties:
\begin{enumerate}
\item A unit of information propagates from one node to another at time $t_{i}$ iff there exists an edge between them at time $t_{i}$.
\item At every time step, each node accumulates all the information it receives.
\end{enumerate}
}

\begin{figure}[tbh]
\begin{tabular}{@{}cc@{}}
  \includegraphics[height=0.63in]{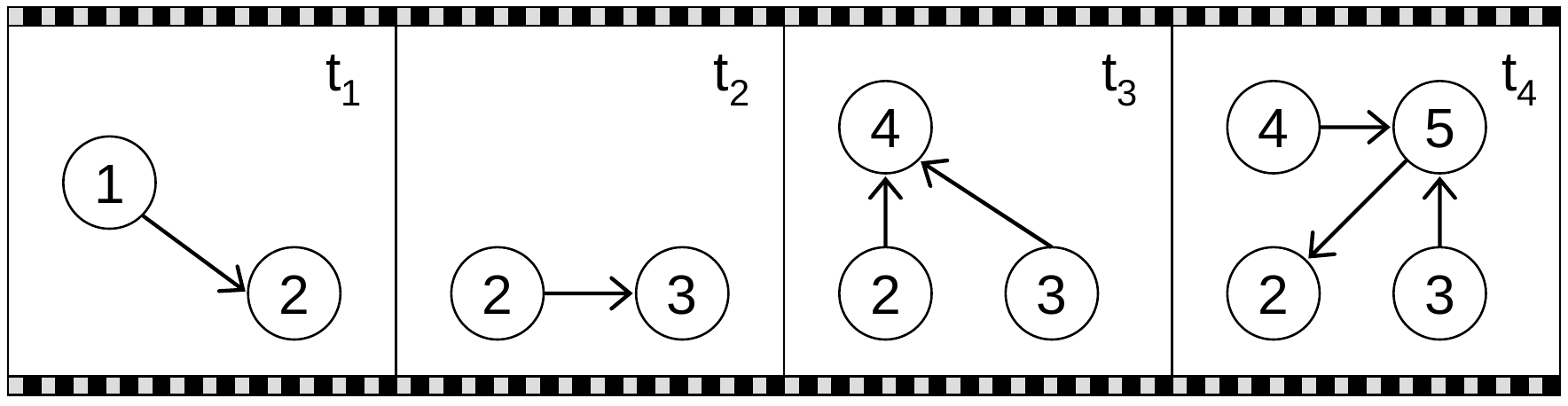} & \includegraphics[height=0.52in]{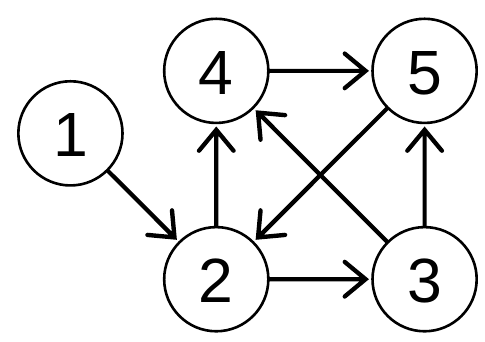}  \\
  (a) & (b)
\end{tabular}
\caption{Example network. (a) Snapshots of the network showing only connected nodes at times $t_1,t_2,t_3$ and $t_4$. (b) A static network that aggregates different snapshots into a single network.}\label{fig:toy}
\end{figure}

Figure~\ref{fig:toy}(a) shows four snapshots of a hypothetical dynamic network, with only connected nodes displayed. Note that edges are directed. A common method to analyze such a dynamic network is to create a static network that aggregates edges observed at all times. Such aggregate network is shown in Fig.~\ref{fig:toy}(b). However, aggregating over all edges loses important temporal information that can help elucidate the structure of a dynamic network~\cite{Braha06}.
Consider how information spreads on a dynamic network. Node $i$ will only be able to send a message to node $j$ at time $t_k$ if and only if there exists an edge between $i$ and $j$ and that time. Specifically, consider how a message sent  by node 1 may reach node 5. In the static network, there are three acyclic paths from 1 to 5: 1$\rightarrow$2$\rightarrow$4$\rightarrow$5, 1$\rightarrow$2$\rightarrow$3$\rightarrow $4$ \rightarrow $5, and 1$ \rightarrow $2$ \rightarrow $3$ \rightarrow $5. Not all these paths are physically realizable, however. If a node does not retain a message but transmits it in the next time step, the only meaningful path is 1$\rightarrow$2$\rightarrow$3$\rightarrow$4$\rightarrow $5. Using this intuition, we define a novel centrality metric for dynamic networks that computes  the number of paths between nodes $i$ to $j$ that exist over a period of time.

\subsection{Memoryless Formulation}
\label{sec:memoryless}
We assume that the future state of the network $G_{t_{k+1}}$ depends only on its current state $G_{t_{k}}$, and none of its past states. This implies that each node propagates information it receives in the current time step at the very next time step. We model information spread on a network as a memoryless dynamic process:
\vspace{-15pt}
\blist
\item with probability $\beta^{t_k}$, a node initiates transmission of information by sending a message to its neighbors at time $t_k$
\item with probability $\alpha_{k}^{t_{k+1}}$, a node sends the message it received at time $t_k$ to its neighbors at time $t_{k+1}$
\elist
\vspace{-15pt}
Although in principle, the attenuation factors $\alpha$ and $\beta$ can change with time and distance from the source, which can be easily modeled in this framework, for simplicity we assume that all  $\alpha_{k}^{t_{i}}=\alpha $ and $\beta^{t_i}=\beta$.
The expected amount of information sent by node $i$ at time $t_1$ that reaches node $j$ at time $t_n$ via a sequence of intermediate nodes is given by the $(i,j)$'s element of the \textit{dynamic centrality matrix}:
\remove{
\begin{eqnarray*}
\label{eq:d_c}
C^{d}_{\Delta_{1,n}}(\beta^{t_1},\alpha_{1}^{t_{2}},\cdots ,\alpha_{n-1}^{t_{n}}) &=&\beta^{t_1}\{A(t_{1})\\
&+&\alpha_{1}^{t_{2}} A(t_{1})A(t_{2})\\
&+&\alpha_{1}^{t_{2}}\alpha_{2}^{t_{3}} A(t_{1})A(t_{2})A(t_{3})\\
&+&\cdots\\
&+& \alpha_{1}^{t_{2}} \cdots \alpha_{n-1}^{t_{n}} A(t_{1})\cdots A(t_{n})\}
\end{eqnarray*}
}
\begin{eqnarray}
\label{eq:d_c}
C^{d}_{t_1 \rightarrow t_n}(\beta,\alpha) &=&\beta A(t_{1}) +\beta \alpha A(t_{1})A(t_{2}) + \cdots \nonumber\\
&&+  \beta \alpha^{n-1} A(t_{1})\cdots A(t_{n})\,.
\end{eqnarray}
Let $\Delta_{1,n}$ be the time interval $\{t_{1},\ldots,t_{n}\}$ that information propagates from any node $i$ at time $t_{k}$ to any node $j$ at time $t_{n}$, $1 \le k \le n$. The cumulative expected amount of information reaching node $j$ from node $i$ in a given time interval $\Delta_{1,n}$ is given by
the $i,j$'s element of the  \textit{cumulative dynamic centrality} matrix:
\begin{equation}
C^{d}(\beta,\alpha,\Delta_{1,n})= \sum_{k=1}^{n}{C^{d}_{t_k\rightarrow t_n}(\beta, \alpha)}.
\end{equation}

\subsection{Formulation with Memory}
In many dynamic networks, the future state of the network $G_{t_{k+1}}$ may depend not only on its current state, but also on (possibly all) its past states $G_{t_{i}} (i<k)$. In a social network, for example, two individuals will remember an interaction they had, even if it happened a long time ago. Since in most situations more recent interactions are more important, we model this by introducing memory decay characterized by the \textit{retention probability} $\gamma\ (0\le \gamma \le 1)$ and \textit{retention length} $m\ (m \in 1, \cdots, n)$.
We model this as dynamic process with the following properties:
\vspace{-15pt}
\blist
\item with probability $\beta$ a node initiates transmission of information by sending a message to its neighbors at time $t_k$
\item with probability $\alpha$ a node passes the message it received at time $t_k$ to its neighbors at time $t_{k+1}$.
\item with probability $\gamma$ a node retains the message it received at time $t_k$ until time $t_{k+1}$.
\elist
\vspace{-15pt}

The \textit{retained adjacency matrix}  $R(t_n)$ at time $t_n$ depends on adjacency matrices at the previous times:
\remove{
\begin{eqnarray*}
& &A(t_{n})+\gamma A(t_{n-1})+\cdots+\gamma^{n-1}A(t_{1}) ~~ if ~ n<m\\
R(t_{n},\gamma)&= &A(t_{n})+\gamma A(t_{n-1})+\cdots  \\
 & & + \gamma^{m-1}A(t_{n-m+1})~~ otherwise
\end{eqnarray*}
}	
\begin{equation*}
R(t_{n},\gamma) =
\begin{cases}
  A(t_{n})+\gamma A(t_{n-1})\cdots+\gamma^{n-1}A(t_{1}),  & \mbox{if }n<m \\
  A(t_{n})+\gamma A(t_{n-1})+\cdots & \\
   ~~~+ \gamma^{m-1}A(t_{n-m+1}), & \mbox{otherwise }
\end{cases}
\end{equation*}

\remove{
and the \textit{normalized retained adjacency matrix}  $NR(t_n)$ as:\\
\begin{eqnarray*}
NR(t_{n})&=& \frac{A(t_{n})+\gamma A(t_{n-1})+\cdots+\gamma^{n-1}A(t_{1})}{1+\gamma+\cdots+\gamma ^{n-1}}(if ~ n < m)\\
&= &\frac{A(t_{n})+\gamma A(t_{n-1})+\cdots+\gamma^{m-1}A(t_{n-m+1})}{1+\gamma+\cdots+\gamma ^{m-1}} ( otherwise)
\end{eqnarray*}
}

Following Section \ref{sec:memoryless}, the \textit{retained dynamic centrality matrix} can then be given as:
\begin{eqnarray}
\label{eq:r_d_c}
RC^{d}_{t_{1}\rightarrow t_{n}}(\beta,\alpha, \gamma) &=&\beta R(t_{1},\gamma) + \beta \alpha R(t_{1},\gamma)R(t_{2},\gamma) \nonumber \\
&+& \beta \alpha^2 R(t_{1},\gamma)R(t_{2},\gamma)R(t_{3},\gamma)+\cdots \nonumber \\
&+& \beta \alpha^{n-1} R(t_{1},\gamma)\cdots R(t_{n},\gamma)
\end{eqnarray}
\noindent and the  \textit{retained cumulative dynamic centrality} matrix over the time interval $\Delta_{1,n}$ as:
\begin{eqnarray}
\label{eq:r_c_d_c}
RC^{d}(\beta,\alpha,\gamma,\Delta_{1,n}) = \sum_{k=1}^{n}{RC^{d}_{t_k\rightarrow t_n}(\beta,\alpha, \gamma)}
\end{eqnarray}

\begin{figure}[tbh]
\begin{tabular}{@{}c@{}}
  \includegraphics[width=\columnwidth]{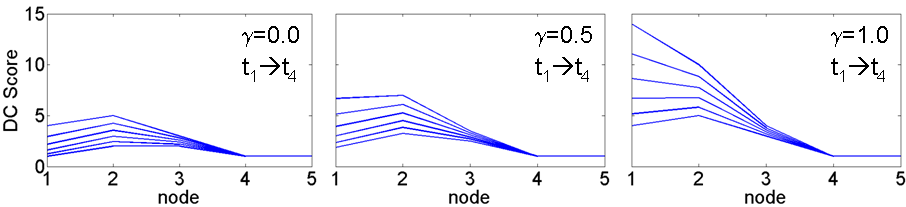}\\
  (a) dynamic \\
    \includegraphics[width=\columnwidth]{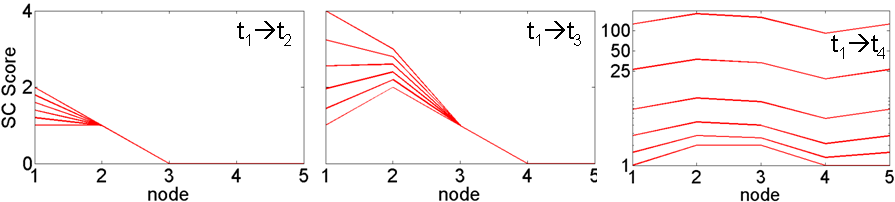}  \\
  (b) static
\end{tabular}
\caption{Dynamic vs static centrality scores for nodes in the dynamic network shown in Fig.~\protect\ref{fig:toy}. (a) Total dynamic centrality scores for different values of $\gamma$ over time period $\Delta_{1,4}$. (b) Total static centrality scores for cumulative networks over time periods $\Delta_{1,2}$, $\Delta_{1,3}$, and $\Delta_{1,4}$. Lines correspond to $\alpha=0.0, 0.2, 0.4, 0.6, 0.8$ and $1.0$ respectively, from the bottom. }\label{fig:toy_rankings}
\end{figure}

\subsection{Ranking in Dynamic Networks}
\label{sec:ranking}
A basic problem in network analysis is ranking nodes to identify important or influential ones. We use dynamic centrality metric to rank nodes in a dynamic network. The intuition behind the ranking scheme is based on the diffusion of information on a network.
Suppose that node $i$ sends a unit of information at time $t_1$. The expected amount of information reaching node $j$ from $i$ over a time interval $\Delta_{1,n}$ is given by $RC^{d}_{ij}(\alpha,\gamma,\Delta_{1,n})$.
\footnote{Since $\beta$ factors out of the equations, without loss of generality we set $\beta=1$.}
The total amount of information sent by $i$ that reaches all other nodes in the network is measured by the \emph{dynamic centrality} of $i$:
\begin{equation}
\label{eq:dc}
DC_i(\alpha,\gamma,\Delta_{1,n})=\sum_{j}RC^d_{ij}(\alpha,\gamma,\Delta_{1,n}).
\end{equation}
\noindent  This metric measures how connected node $i$ is to other nodes in the network over some period of time $\Delta_{1,n}$. Ranking nodes by how well connected they are allows us to identify the most \emph{influential} nodes in a dynamic network over a period of time.

Dynamic programming can be used to efficiently compute dynamic centrality. As can be seen in algorithm \ref{alg:1}, in each iteration, $r_{i}$  depends only on $r_{i-1}$ and $R(t_{n-i},\gamma)$. Since the network at time $t_{i} (i \in 1,\cdots, n)$ is given by graph $G_{t_{i}}= (V_{t_{i}},E_{t_{i}} )$, in the naive implementation of this algorithm, taking $|E|=|\cup_{i}E_{t_{i}}|$  and $|V|=|\cup_{i}V_{t_{i}}|$, each iteration has a runtime complexity of $O(|E|)$ and space complexity of $O(|V|+|E|)$. Assuming that the main memory is just large enough to hold $r_{i}$, $r_{i-1}$  and $DC(\alpha,\gamma,\Delta_{1,n})$, the i/o cost for each iteration is $O(|E|)$.  If main memory is large enough to hold only $r_{i}$, and assuming efficient data structure such as a sorted link list is used to store $R(t_{n-i},\gamma)$, i/o cost is $O(|V|+|E|)$. Since this formulation of dynamic centrality is very similar to that of PageRank~\cite{PageRank}, similar block based strategies can be used to further improve speed and efficiency of computing dynamic centrality \cite{Haveliwala:1999} \cite{Kamvar:2003}. Like PageRank, dynamic centrality can  be implemented using the map-reduce paradigm \cite{Dean:2008}, guaranteeing the scalability of this algorithm and its applicability to very large datasets.

\begin{algorithm}
\caption{ Dynamic centrality}
\label{alg:1}
\begin{algorithmic}
\STATE{\bf{Input}}\\
$\{R(t_k,\gamma): \forall k \in1,2\cdots n\}$: Retained adjacency matrices  \\
$\alpha,\beta$: attenuation factors\\
$e$:unit vector ($ n\times 1$)
\STATE{\bf{Output}}\\
$DC(\alpha,\gamma,\Delta_{1,n})$: Dynamic centrality vector
\STATE  {\bf{Initialize }}\\
 $r_{0} \leftarrow  \beta R(t_{n},\gamma)e$  \\
 $DC(\alpha,\gamma,\Delta_{1,n}) \leftarrow r_{0}$ \\
 \FOR{$i = 1$ to $n-1$}
\STATE $r_{i}  \leftarrow  R(t_{n-i},\gamma)(\beta e+\alpha r_{i-1}) $\\
 $DC(\alpha,\gamma,\Delta_{1,n}) \leftarrow DC(\alpha,\gamma,\Delta_{1,n}) +r_{i}$
\ENDFOR\\
\end{algorithmic}
\end{algorithm}

In addition to ranking nodes, dynamic centrality can be used to identify nodes that have the most influence on a given node over some period of time, or have been most influenced by it. For example, to find the node that is most influenced by $i$, we identify node $j$ with the largest value of $RC^{d}_{ij}$, given by Eq.~\ref{eq:r_c_d_c}. Similarly, $RC^{d}_{ji}$ gives the influence of node $j$ on $i$ and can be used to identify nodes that have had the most influence on $i$ over some period of time.

Tunable parameters $\alpha$ and $\gamma$ enable us to use dynamic centrality to study the structure of dynamic networks at different time and length scales.
As described in Section \ref{sec:related}, $\alpha$ sets the length scale of interactions. As $\alpha$ grows, longer paths become more important, and dynamic centrality takes into account increasingly larger network components. Parameter $\gamma$ sets the time scale of the interactions. For $\gamma=0.0$, only the most recent interactions are taken into account. As $\gamma$ grows, older interactions are also considered. In the extreme case of perfect retention or memory, $\gamma=1.0$, every past interaction is remembered, similar to how a cumulative version of a dynamic network is constructed.

We apply dynamic centrality to study the toy network shown in Fig.~\ref{fig:toy}(a).
Figure~\ref{fig:toy_rankings} plots dynamic centrality score of each node, which is given by $DC_i(\alpha,\gamma,\Delta_{1,4})$. Each plot shows results for a different value of $\gamma$, and each line in the plot corresponds to a different value of  $\alpha$ from $0.0$ to $1.0$ in steps of $0.2$ from the bottom.
For $\gamma \le 0.5$ node 2 has the highest score for all values of $\alpha$, and is therefore, highest ranked, although for $\alpha=0.0$, $\gamma=0.0$ node 3 has the same $DC$ score as node 2. While both 2 and 3 have two outgoing edges, a larger number of longer paths originate from node 2 (2$\rightarrow$3$\rightarrow$4$\rightarrow$5, 2$\rightarrow$4$\rightarrow$5, 2$\rightarrow$3$\rightarrow$5) than node 3 (3$\rightarrow$4$\rightarrow$5, 3$\rightarrow$5). In the case of perfect memory ($\gamma=1.0$), node 2 is the highest ranked node for $\alpha \le 0.4$.
As longer paths become more important at larger values of $\alpha$, node 1's influence grows and it becomes highest ranked. As the earliest node to send a message, it is the origin of the longest paths in the network.

We compare dynamic centrality-based rankings with those produced by an equivalent static metric that computes the number of attenuated paths in an aggregate network shown in Fig.~\ref{fig:toy}(b) regardless of the time the links were formed. To compute the static centrality score, we use $C_i^s(\alpha)=\sum_j{C^s_{ij}(\alpha)}$, where $C^s_{ij}(\alpha)$ is given by Eq.~\ref{eq:static}. Figure~\ref{fig:toy_rankings}(b) shows static centrality scores for cumulative network that aggregate edges over time periods $\Delta_{1,2}$, $\Delta_{1,3}$, and $\Delta_{1,4}$. The aggregate network corresponding to the period $\Delta_{1,4}$  is shown in Fig.~\ref{fig:toy}(b). Static centrality leads to a radically different ranking.
In the static networks that aggregate edges over periods $\Delta_{1,2}$ and $\Delta_{1,3}$, node 1 is considered most influential, except for small values of $\alpha$ in the middle plot, when node 2 becomes more influential.
Because of cycles introduced at the last time step (by 5$\rightarrow$2 edge), the static centrality scores computed for the network aggregated over the period $\Delta_{t,4}$ (last plot in Fig.~\ref{fig:toy_rankings}(b)) grow large with $\alpha$.\footnote{We keep the first 10 terms in the sum in Eq.~\protect\ref{eq:static}. This keeps $C^s$ from growing too large.} Node 2 is most important for all values of $\alpha$, followed closely by nodes 1 and 3. Surprisingly, node 5 is judged to be very influential, surpassing node 4 in score. This is obviously wrong, since only a single path of length one originates from node 5 in the dynamic network.

\begin{figure}[tbh]
\begin{tabular}{@{}c@{}}
  \includegraphics[width=1\columnwidth]{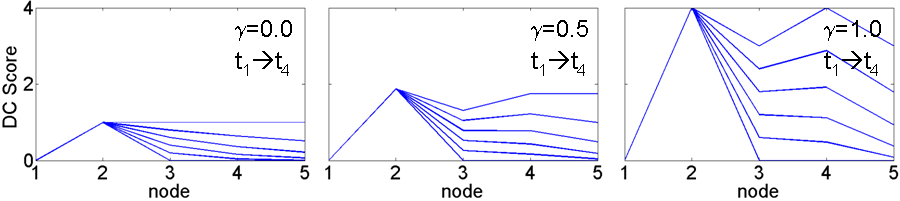}\\
  (a) dynamic \\
    \includegraphics[width=1\columnwidth]{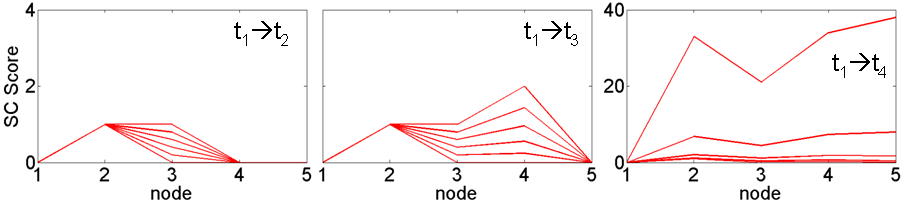}  \\
  (b) static
\end{tabular}
\caption{Influence of node 1 on others in the dynamic vs static centrality formulations. (a) Dynamic influence of node 1 on others in the dynamic network for different values of $\gamma$ over time period $\Delta_{1,4}$. (b) Static influence of node 1 on others in cumulative networks over time periods $\Delta_{1,2}$, $\Delta_{1,3}$ and $\Delta_{1,4}$. Lines correspond to $\alpha=0.0, 0.2, 0.4, 0.6, 0.8$ and $1.0$ respectively, from the bottom. }\label{fig:toy_influence}
\end{figure}

In addition to ranking nodes, we can look at a given node's influence \emph{on} other nodes in the network. Figure~\ref{fig:toy_influence} shows the influence of node 1 computed using Eq.~\ref{eq:static} and Eq.~\ref{eq:r_c_d_c}, for different values of $\alpha$ and $\gamma$. Again, the static and dynamic formulations lead to different views of influence. Dynamic centrality metric finds that node 1 has most influence on node 2, although as $\alpha$ and $\gamma$ increase, its influence on node 4 grows to be comparable to its influence on node 2. This is reasonable, because since node 1 is directly connected to 2, we expect it to have most influence on that node. Node 4 is connected to node 1 through nodes 2 and 3, and will also be highly influenced by it. Although node 5 is also linked to 1 by multiple paths, these paths are longer than those connecting node 1 to 3; therefore, node 1's influence on 5 should be less than on 4. However, the static centrality metric {applied to the aggregate network} finds that node 1 has biggest influence on node 5, followed by 4 and 2. Even when links are aggregated over a shorter period, $\Delta_{1,3}$, node 4 is most influenced by 1 at larger values of $\alpha$.

In summary, static and dynamic formulations of centrality lead to widely different views of importance in a dynamic network. We claim that by taking into account constraints on information flow imposed by the temporal ordering of edges, dynamic centrality formulation leads to a more accurate understanding of the structure of dynamic networks.

\section{Citations Network}
\label{sec:citations}
The citations data set consists of articles uploaded to the theoretical high energy physics (hep-th) section of the \emph{arXiv} preprints server from 1993 to April, 2003.\footnote{www.cs.cornell.edu/projects/kddcup/datasets.html} There are about 28,000 articles with about 350,000 citations. Each article is identified by a unique number, with first two digits representing the year of submission. Data was cleaned by removing citations to articles that appeared in the future, as well as citations of the article to itself.

We partition the data by year to construct snapshots of the dynamic network in consecutive years.
The citations made by papers uploaded to \emph{arXiv} during some year form the edges of the snapshot for that year.  A year may not be an optimal partition of the data, since a small number of articles published in one year cite others published in the same year, but it is a convenient time scale to measure scientific production and interaction between researchers. We transpose the adjacency matrix to reverse direction of edges so that it represents the flow of influence from cited to citing articles.  Citations data can be alternately represented by a static network that aggregates all edges that appear over some time period, e.g., 1993--2003. Several researchers analyzed the structure of the static aggregate network, e.g., with PageRank algorithm, to identify influential articles~\cite{Redner05,Chen07,Walker06,Sayyadi09sdm}. In contrast, we explicitly take the dynamic nature of the network into account.

\begin{figure}[htbp]
\begin{tabular}{@{}cc@{}}
\includegraphics[width=0.47\linewidth]{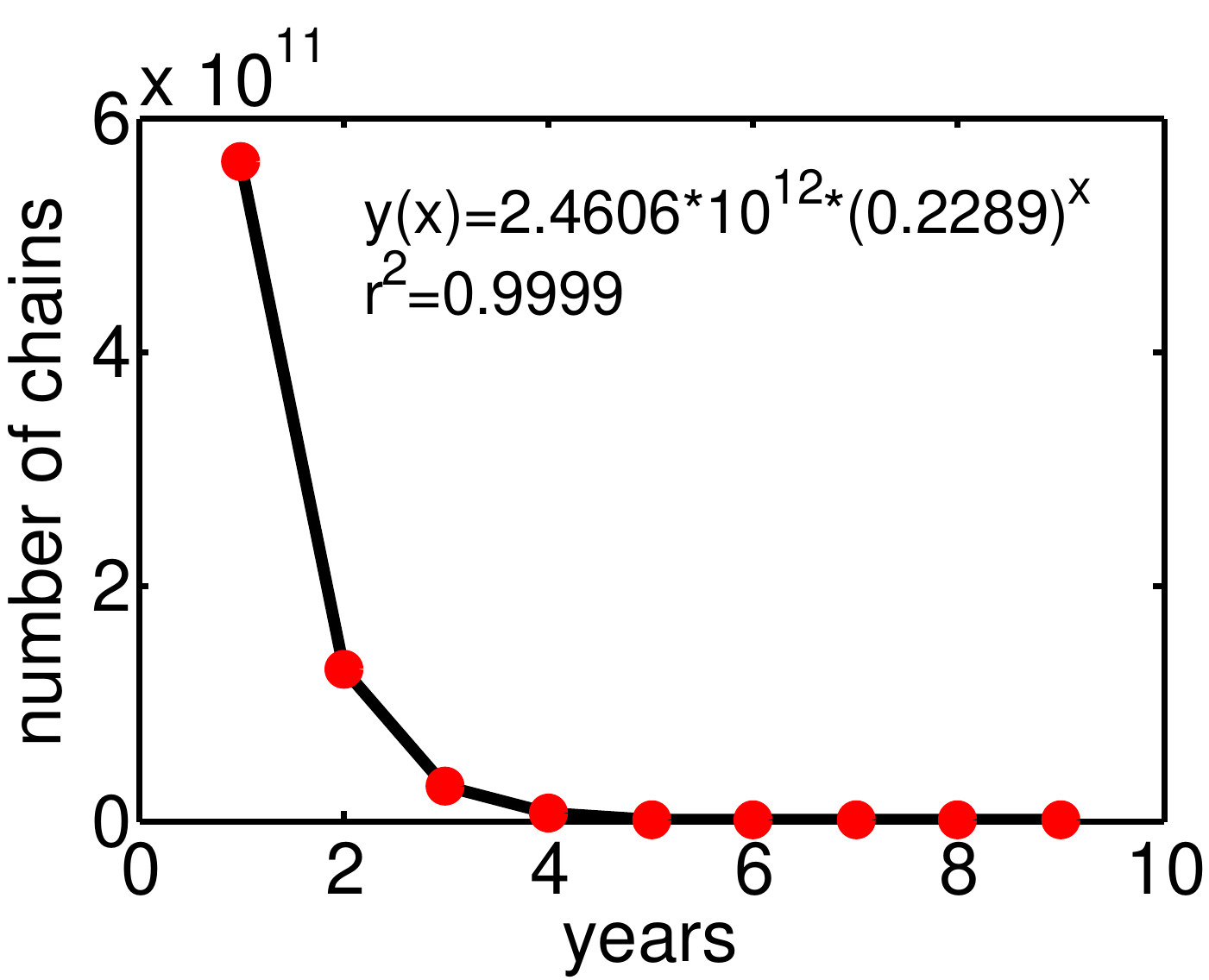}  &
\includegraphics[width=0.48\linewidth]{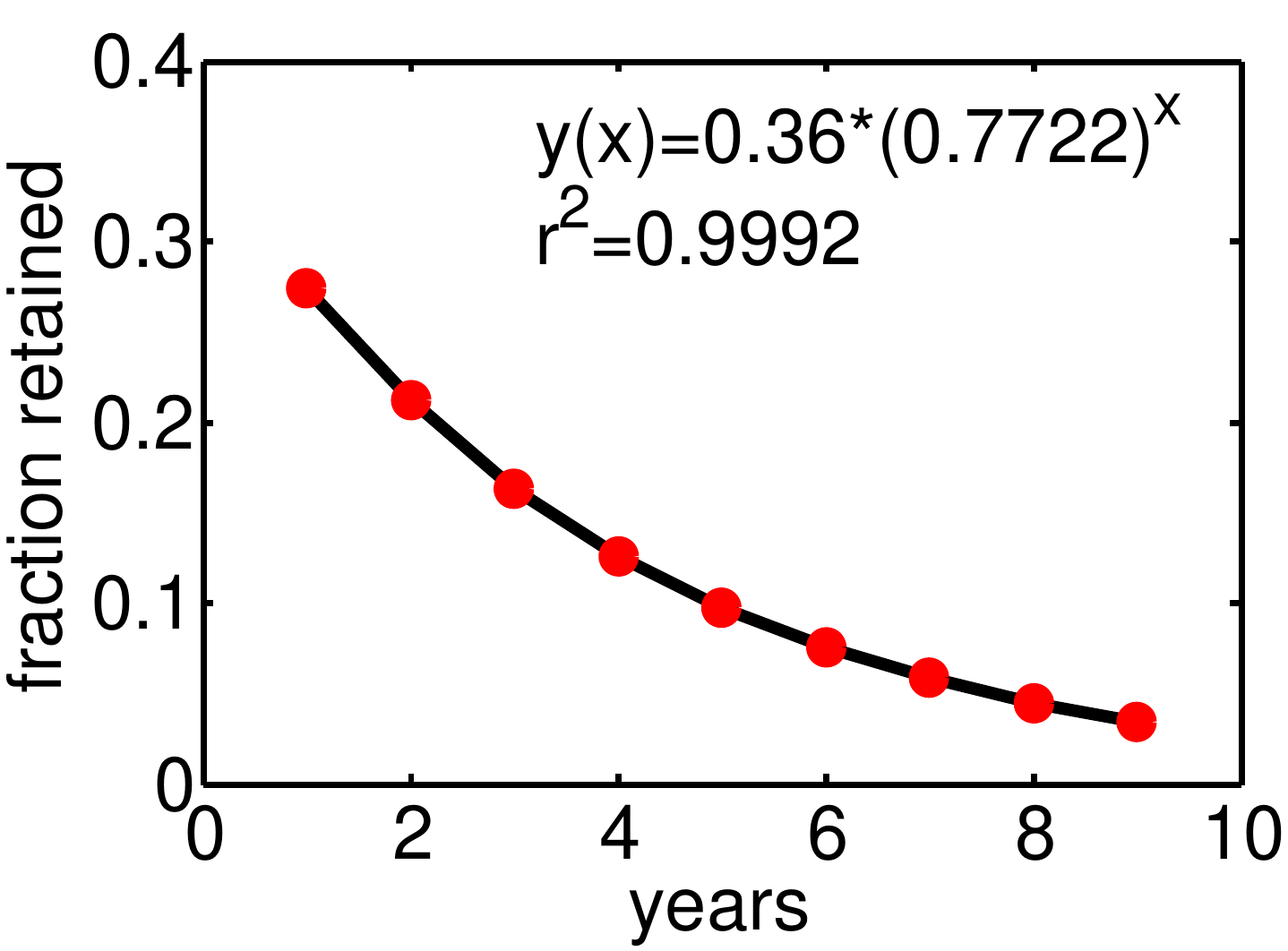}\\
(a) & (b)
\end{tabular}
\caption{Parameter estimation for the \emph{arXiv} data set. (a) Distribution of the number of citations chains of different length with fit. (b) Distribution of the fraction of citations to papers published $x$ years previously with fit.}
\label{fig:fits}
\end{figure}

\subsection{Parameter Estimation}
\label{sec:parameters}
Dynamic centrality metric contains parameters $\alpha$ and $\gamma$. While varying their values turns dynamic centrality into a tool to study the structure of the network at different time and length scales, a natural question is what are the appropriate values for these parameters? If we have enough data about the network, we can estimate them directly from the data. In this section we describe the methodology to estimate optimal values of $\alpha$ and $\gamma$ for the \emph{ArXiv} data.

To estimate $\alpha$, we find the distribution of citation chains that span consecutive years. In other words, we set $\gamma=0$, so that no older citations are retained. $N_j$ gives the total number of chains of length $j$ that start in year $t_{n-j+1}$ and end in year $t_n$. Assuming that the probability of picking a chain is proportional to the probability of transmitting a message along the chain, $N_j$ decays geometrically with $\alpha$. Therefore, the probability of choosing a citations chain of length $j$ is given by $\alpha^{j}$. The expected number of citation chains is $E(N_{j})=\alpha E(N_{j-1}).$ Figure~\ref{fig:fits}(a) plots the distribution of the number of chains in the \emph{ArXiv}  data set that end in the year $t_n=2002$. This distribution is well fit (with $R^2=0.9999$) by $E(N_{j})=c \cdot 0.2289^{j}$, where $c=2.4606 \times 10^{12}$. This gives us  $\alpha=0.2289$ for the \emph{arXiv} data set. At this value of $\alpha$, the mean path has length ${1}/(1-\alpha)=1.3$. This is consistent with the observation that citations chains have length $\simeq 2$ \cite{Walker06,Chen07}.

To estimate $\gamma$, we assume that citation retention probability decays geometrically with time~\cite{Redner05}.  Let $C^{j}_{k}$  be the number of papers at time $j-k$ cited by papers at time $j$. Since the number of citations increases in time, we calculate $W^{j}_{k}= C^{j}_{k}/\sum_{k} C^{j}_{k}$, the fraction of papers appearing at time $j-k$ that are cited by papers at time $j$. Taking the average of $W^{j}_{k}$ for all $j$, gives the expected fraction of citations in a given paper to papers published $k$ years before it, $E(W_{k})$. Therefore according  to our hypothesis, $E(W_{k}) =\gamma E(W_{k-1})$. Figure~\ref{fig:fits}(b) plots this distribution for papers in the \emph{arXiv} data set. Data is well fit ($R^2=0.9992$) by $E(W_{k})=d \cdot (0.7722)^{k}$, where $d=0.36$. Hence, $\gamma= 0.7722$.

\begin{figure}[htb]
\begin{tabular}{@{}cc@{}}
  \setlength{\tabcolsep}{1pt}
\includegraphics[width=0.48\linewidth]{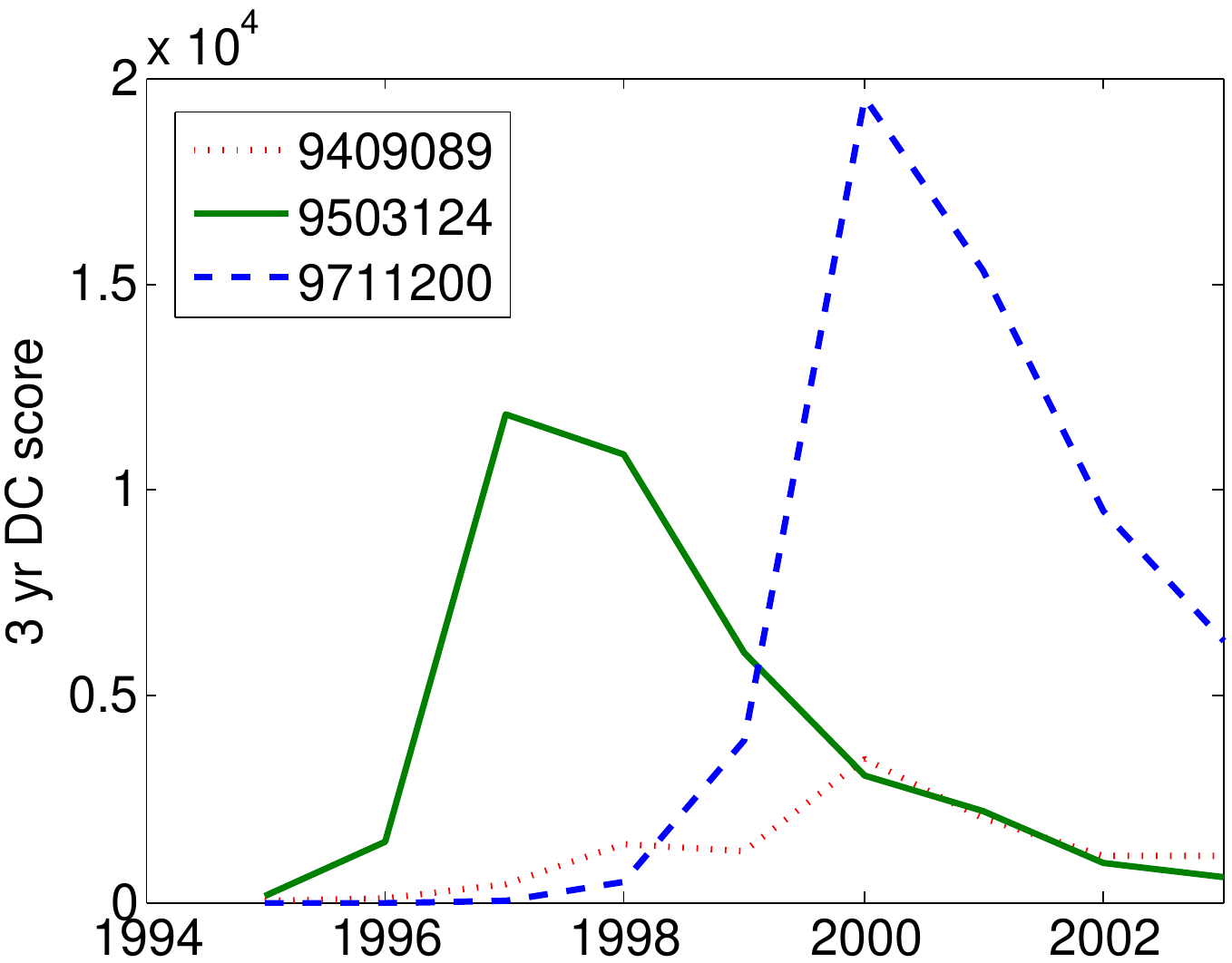}  &
\includegraphics[width=0.48\linewidth]{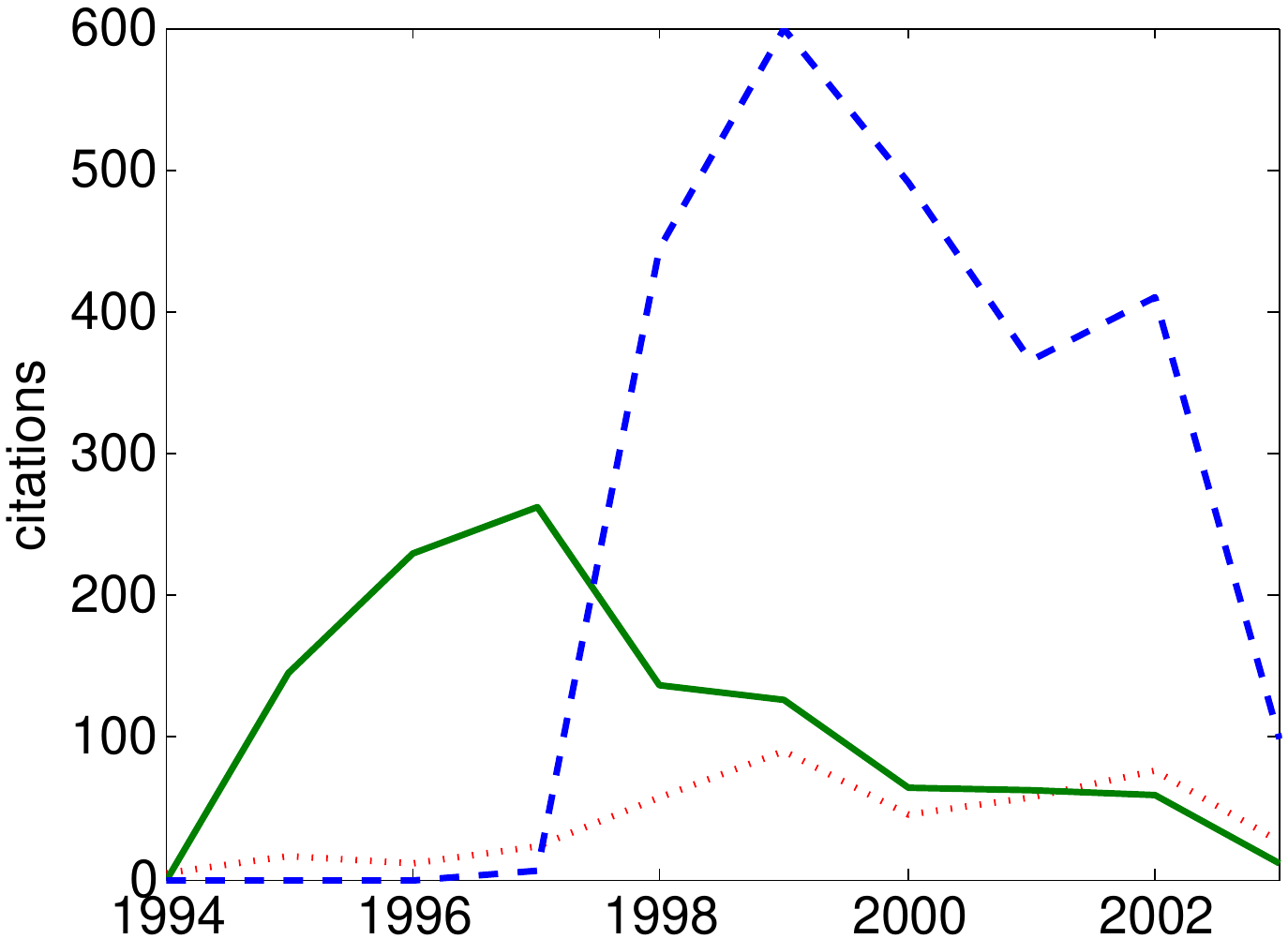} \\
(a) & (b)
\end{tabular}
\caption{Evolution of influence of three articles. (a) Dynamic centrality scores computed over a rolling three year window vs time. (b) Number of citations received by papers each year vs time.}
\label{fig:3yearDC}
\end{figure}

\subsection{Influence of Individual Articles}

Dynamic centrality, Eq.~\ref{eq:dc}, provides insights into evolution of scientific topics and influence of individual articles. Figure~\ref{fig:3yearDC}(a) shows how DC scores of three articles change in time. These articles were randomly chosen from among the articles ranked highest by PageRank.
DC scores of the three articles were computed over a sliding three year window using optimal parameters $\alpha=0.2289$ and $\gamma=0.7722$. This time window means that the longest citations chains DC will consider are of length three. Since there is evidence that researchers do not often follow citations links more than two levels deep~\cite{Walker06,Chen07}, a window of size three will adequately capture longer range interactions in this network. Evolution of article's centrality (Fig.~\ref{fig:3yearDC}(a)) shows a similar trend to the number of new citations it receives each year (Fig.~\ref{fig:3yearDC}(b)).
\remove
{
Both measures rise and gradually decline, though influence grows faster, as can be clearly seen for article 9503124. Article 9409089 has two peaks in influence, in 1998 and 2000, with the second peak higher than the first. Its popularity, as measured by citations rate, is also double-peaked, with the first peak somewhat higher than the second. There is a dip in popularity in 2000, which actually corresponds to the article's highest influence.
}

\begin{table*}[tbh]
  \centering
  \setlength{\tabcolsep}{1pt}
  \scalebox{0.8}
{
  \begin{tabular}{|c|p{68mm}|c||c|p{58mm}|c||c|p{45mm}|c|}
    \hline
\multicolumn{3}{|c||}{\scriptsize{9409089}} & \multicolumn{3}{|c||}{\scriptsize{9503124}} & \multicolumn{3}{|c|}{\scriptsize{9711200}} \\ \hline
 \multicolumn{2}{|c|}{\scriptsize{influenced by}} 	 &	\scriptsize{cites}	&  \multicolumn{2}{|c|}{\scriptsize{influenced by}}   	& \scriptsize{cites}	&	 \multicolumn{2}{|c|}{\scriptsize{influenced by}} 	& \scriptsize{cites}	\\ \hline

\scriptsize{9311037} &	\scriptsize{High Energy Asymptotics of Multi--Colour QCD} &	1 &	\scriptsize{9207053} &	 \scriptsize{Electric Magnetic Duality in Str. Th.} &	0	& \scriptsize{9207053} &	\scriptsize{Electric Magnetic Duality in Str. Th.} &	0 \\

\scriptsize{9308139} &	\scriptsize{Strings, Black Holes and Lorentz Contraction} &	1 &	\scriptsize{9211056} &	 \scriptsize{Magnetic Monopoles in Str.Th.} &	0&		\scriptsize{9205027	} &	\scriptsize{Supersymmetry as a Cosmic Censor} &	0\\

\scriptsize{9402125} &	\scriptsize{String Thermalization at a Black Hole Horizon} &	1&		\scriptsize{9209016} &	 \scriptsize{Electric-Magnetic Duality $\ldots$} &	0&		\scriptsize{9207016} &	\scriptsize{ Noncompact Symmetries in Str. Th.} &	0 \\

\scriptsize{9306069} &	\scriptsize{The Stretched Horizon and Black Hole Complementarity} &	0	& \scriptsize{9402002} &	 \scriptsize{ Strong-Weak Coupling Duality in 4D Str. Th.} &	1 &	\scriptsize{9211056} &	\scriptsize{Magnetic Monopoles in Str.Th.} &	0 \\

\scriptsize{9307168} &	\scriptsize{String Theory and the Principle of Black Hole Complementarity} &	0	&	 \scriptsize{9208055} &	\scriptsize{Putting String/Fivebrane Duality to the Test} &	0	&	 \scriptsize{9305185} &	 \scriptsize{ Duality Symmetries of 4D Heterotic Strings} &	0 \\

\scriptsize{9308100} &	\scriptsize{Gedanken Experiments involving Black Holes} &	0&	 	\scriptsize{9207016} &	 \scriptsize{ Noncompact Symmetries in String Th.} &	0&	 	\scriptsize{9209016} &	\scriptsize{Electric-Magnetic Duality $\ldots$} & 0 \\

\scriptsize{9204002} &	\scriptsize{Classical and Quantum Considerations of 2d Gravity} &	0&	 	\scriptsize{9205027} &	\scriptsize{Supersymmetry as a Cosmic Censor} &	 1&	 	\scriptsize{9208055} &	\scriptsize{Putting String/Fivebrane Duality to the Test} &	0 \\

\scriptsize{9201061} &	\scriptsize{Are Horned Particles the Climax of Hawking Evaporation?	} & 0&	 	 \scriptsize{9303057} &	\scriptsize{Magnetic Monopoles} &	0	&	 \scriptsize{9304154} &	\scriptsize{Duality Symmetric Actions} &	0 \\

\scriptsize{9201074} &	\scriptsize{Black Hole Evaporation in 1+1 Dimensions} &	0	&	 \scriptsize{9304154} &	 \scriptsize{Duality Symmetric Actions} &	0	&	 \scriptsize{9303057} &	\scriptsize{Magnetic Monopoles} &	0 \\

\scriptsize{9207034} &	\scriptsize{Quantum Theories of Dilaton Gravity} &	0&	 	\scriptsize{9407087} &	 \scriptsize{Monopole Condensation, And Confinement In N=2 Supersymmetric Yang-Mills Theory} &	1	&	 \scriptsize{9410167} &	\scriptsize{Unity of Superstring Dualities} &	0 \\

 \hline
 \end{tabular}
 }
\caption{Ten articles that had the most influence on each of the three target articles computed at optimal $\alpha$ and $\gamma$. Cites column has ``1'' if the target article cites the listed article. Titles of target articles are: ``The World as a Hologram'' (9409089), ``String Theory Dynamics In Various Dimensions'' (9503124), and 	``The Large N Limit of Superconformal Field Theories and Supergravity'' (9711200).
}
\label{tbl:influence}
\end{table*}

In addition to ranking articles, dynamic centrality allows us to directly measure the influence of one article \emph{on} another.  An article will often directly cite another that influenced it. At other times, however, we can trace the history of intellectual contribution through the chain of citations even in the absence of direct citation. The more citations chains link an article to a given article, the more influential the former will be. Table~\ref{tbl:influence} lists the articles found to have the biggest influence on the three articles in figure.~\ref{fig:3yearDC}. Only a fraction of these articles are directly cited by the three target articles. Article 9409089 (by L.~Susskind) deals with the relationship between string theory and black holes. This appears to be a highly specialized topic.  Five of the ten articles found to have most influence on 9409089 were authored by Susskind and collaborators. Articles  9503124 (by E. Whitten) and 9711200 (by J. Maldacena) deal with the more general topic of mathematics of string theory. There is significant overlap in the topics of these papers, as manifested by overlap in the influencing articles. Interestingly, five of the most influential articles (9207053, 9209016, 9402002, 9303057, 9304154) were authored by A.~Sen, pointing to that authors importance in the field.
Although we do not report it, it is interesting to see the papers that were most influenced \emph{by} the target papers. All three target papers highly influenced articles on Supersymmetry, supergravity, holographic renormalization, and AdS/CFT correspondence. Articles 9503124 and 9711200 also influenced papers dealing with ``branes'', a popular subfield of string theory that emerged in the late 1990's.

While it is difficult for a non-specialist to fully evaluate these results,
they appear to be significant. It is highly unlikely the list of papers that highly influenced 9409089 would fortuitously include so many papers dealing black holes and gravity. Likewise, non-existence of magnetic monopoles violates electric-magnetic symmetry, or duality, which has apparently attracted much speculation by string theorists. Appearance of so many papers dealing with these topics in the list of papers that influenced 9503124 and 9711200 cannot by coincidental. These observations give us confidence that dynamic centrality discovers significant relations in the data.

\begin{figure}[tbh]
\begin{tabular}{@{}cc@{}}
  \includegraphics[width=0.48\columnwidth]{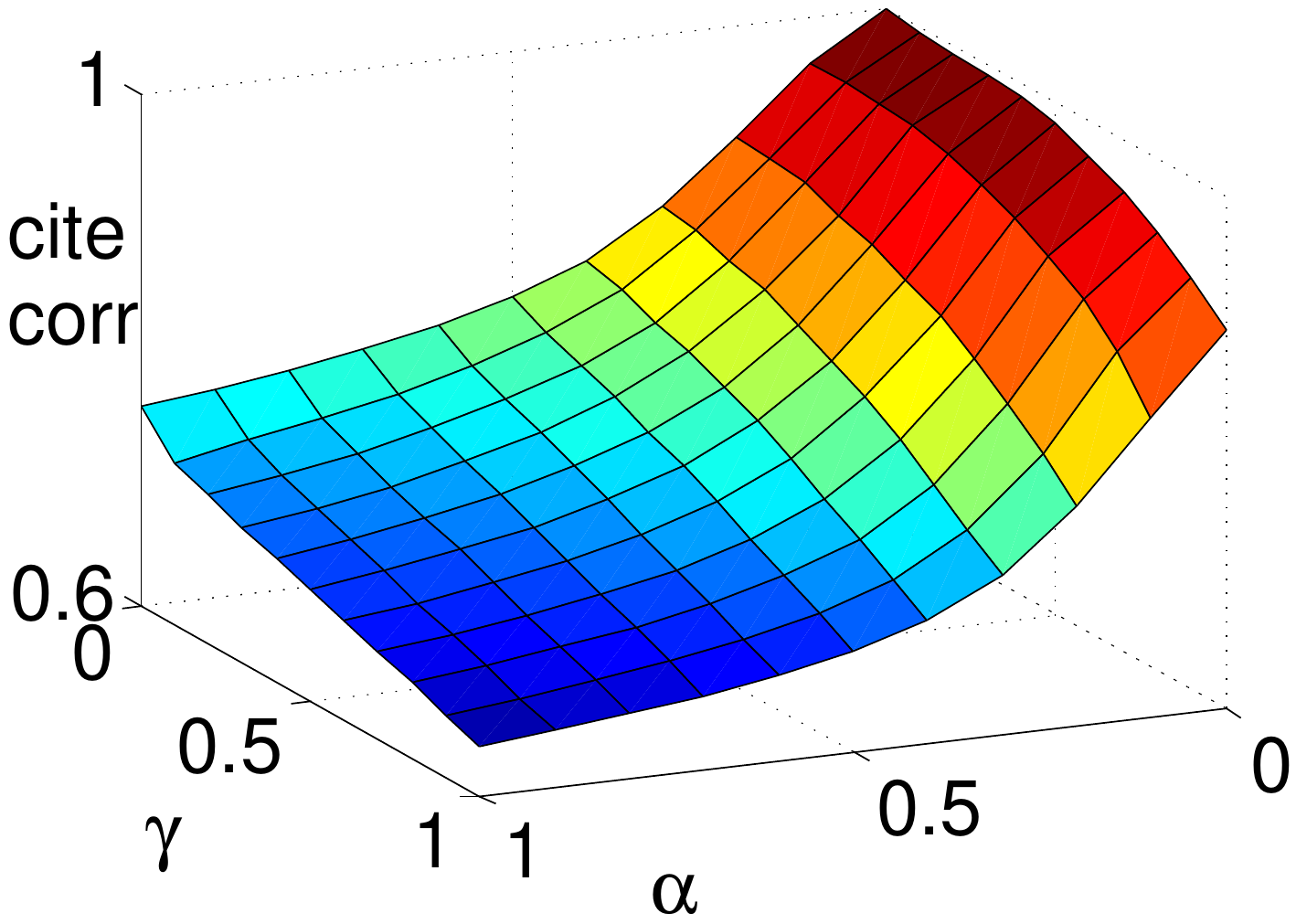} &
    \includegraphics[width=0.48\columnwidth]{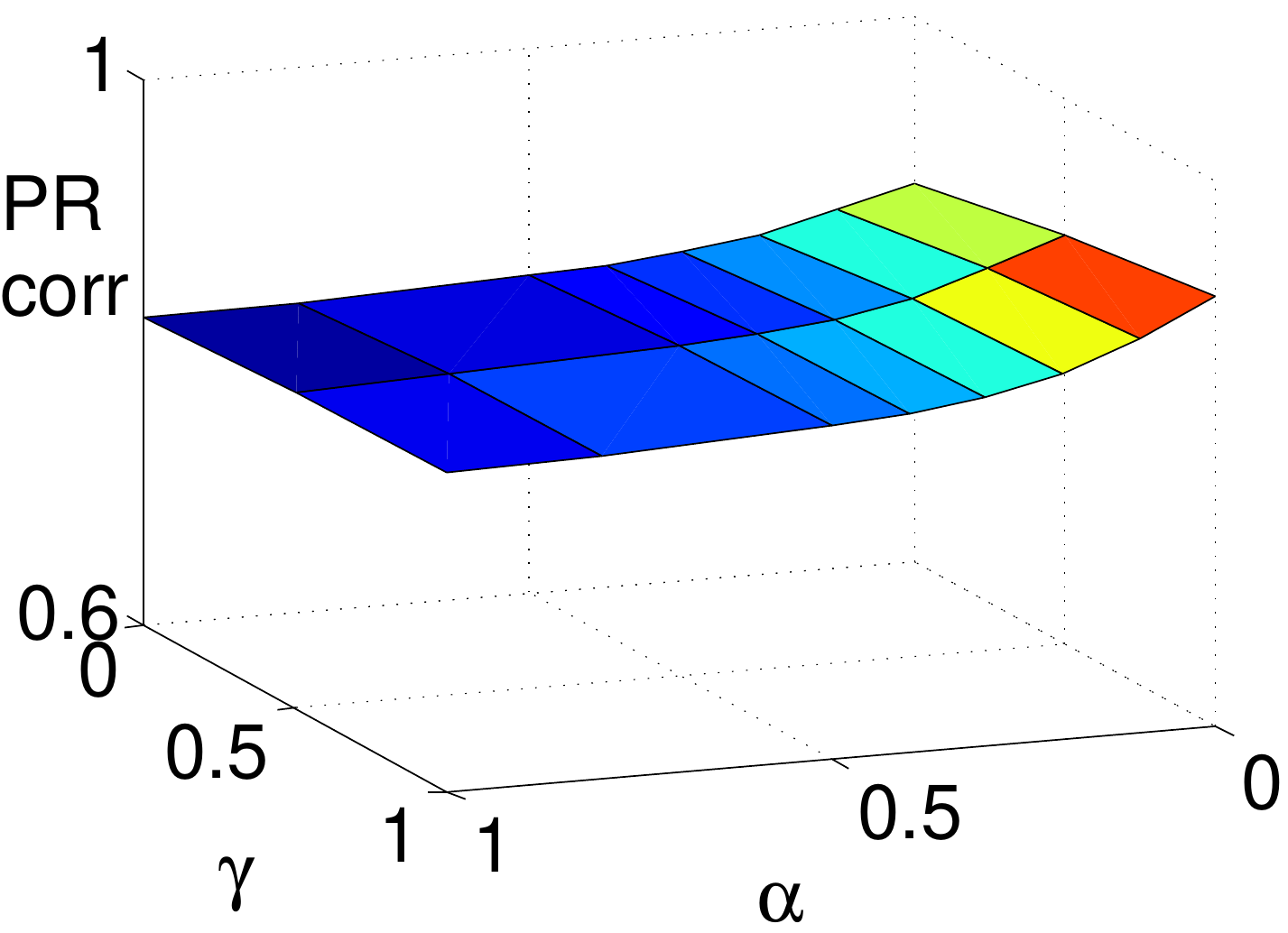}
\end{tabular}
\caption{Spearman's correlation between dynamic centrality-based rankings over the period 1993 -- 2000 and rankings based on articles' total citations count and  PageRank over the same time period. }\label{fig:cite_corr}
\end{figure}

\subsection{Overall Influence and Ranking}
In addition to its usefulness in studying trends in citations data, we can also use dynamic centrality to compute the overall influence of articles over some period and rank them accordingly. This is a common task in bibliometric analysis. While many  metrics have been developed to address this problem, most familiar ones are citations count and PageRank.
Figure~\ref{fig:cite_corr} shows Spearman's rank correlation coefficient between DC rankings and rankings based on total citations count and PageRank. All metrics were computed for the period 1993--2000 inclusively. DC rankings are best correlated with total citations count for $\alpha=0$, $\gamma=0$. This is reasonable, since at these parameter values only direct edges (i.e., citations) contribute to DC. Correlation decreases with both $\alpha$ and $\gamma$, as longer paths and memory are taken into account. For $\alpha \sim 1$, $\gamma \sim 1$ DC rankings are very different from those based on citations count. Correlation with PageRank,\footnote{We used 0.1 as the probability of a random jump in our implementation of the PageRank algorithm.} on the other hand, which was computed on the aggregate static network, is highest for $\alpha=0$, $\gamma=1$. Again, this is expected, since for these parameter values dynamic network  resembles the static network. Correlation with PageRank is worst for $\alpha=1$, $\gamma=0$, i.e., when paths of all length are taken into account and past citations are not retained.

{
\begin{table*}[tbh]
  \centering
  \setlength{\tabcolsep}{1pt}
  \scalebox{0.9}
{
  \begin{tabular}{|c|c|p{90mm}|c|c||c|p{60mm}|c|c|}
    \hline
 & \multicolumn{4}{|c||}{\scriptsize{$\alpha=0$}, \scriptsize{$\gamma=0$}} & \multicolumn{4}{|c|}{\scriptsize{$\alpha=0.2$}, \scriptsize{$\gamma=0$}} \\ \hline
 \scriptsize{DC} &	\scriptsize{arxiv id} &	\scriptsize{title} &	\scriptsize{\#C} &	\scriptsize{PR}  &	 \scriptsize{arxiv id} &	\scriptsize{title} &	\scriptsize{\#C} &	\scriptsize{PR} \\ \hline
1 &	\scriptsize{9711200} &	\scriptsize{The Large N Limit of Superconformal Field Theories and Supergravity	} &	 \scriptsize{2414} &	\scriptsize{6} &	\scriptsize{9503124} &	\scriptsize{String Theory Dynamics In Various Dimensions} &	\scriptsize{1114} &	\scriptsize{2} \\
2 &	\scriptsize{9802150} &	\scriptsize{Anti De Sitter Space And Holography} &	\scriptsize{1775} &	\scriptsize{16} &	 \scriptsize{9410167} &	\scriptsize{Unity of Superstring Dualities} &	\scriptsize{748} &	\scriptsize{5} \\
3 &	\scriptsize{9802109} &	\scriptsize{Gauge Theory Correlators from Non-Critical String Theory} &	\scriptsize{1641} &	 \scriptsize{19} &	\scriptsize{9510017} &	\scriptsize{Dirichlet-Branes and Ramond-Ramond Charges} &	 \scriptsize{1155} &	\scriptsize{3} \\
4 &	\scriptsize{9407087} &	\scriptsize{Monopole Condensation, $\ldots$ Supersymmetric Yang-Mills Theory} &	 \scriptsize{1299} &	\scriptsize{1} &	\scriptsize{9207053} &	\scriptsize{Electric Magnetic Duality in String Theory} &	\scriptsize{102} &	\scriptsize{20} \\
5 &	\scriptsize{9610043} &	\scriptsize{M Theory As A Matrix Model: A Conjecture} &	\scriptsize{1199} &	\scriptsize{9} &	 \scriptsize{9205027} &	\scriptsize{Supersymmetry as a Cosmic Censor} &	\scriptsize{191} &	\scriptsize{10} \\
6 &	\scriptsize{9510017} &	\scriptsize{Dirichlet-Branes and Ramond-Ramond Charges} &	\scriptsize{1155} &	 \scriptsize{3} &	\scriptsize{9207016} &	\scriptsize{Noncompact Symmetries in String Theory} &	\scriptsize{218} &	 \scriptsize{31} \\
7 &	\scriptsize{9908142} & \scriptsize{String Theory and Noncommutative Geometry} &	\scriptsize{1142} &	\scriptsize{47} &	\scriptsize{9305185} &	\scriptsize{Duality Symmetries of 4D Heterotic Strings} &	\scriptsize{171} &	 \scriptsize{14} \\
8 &	\scriptsize{9503124} &	\scriptsize{String Theory Dynamics In Various Dimensions} &	\scriptsize{1114} &	 \scriptsize{2} &	\scriptsize{9211056} &	\scriptsize{Magnetic Monopoles in String Theory} &	\scriptsize{68} &	 \scriptsize{25} \\
9 &	\scriptsize{9906064} &	\scriptsize{An Alternative to Compactification} &	\scriptsize{1030} &	\scriptsize{35} &	 \scriptsize{9510135} &	\scriptsize{Bound States Of Strings And $p$-Branes} &	\scriptsize{775} &	\scriptsize{12} \\
10 &	\scriptsize{9408099} &	\scriptsize{Monopoles, Duality and Chiral Symmetry Breaking in N=2 Supersymmetric QCD} &	 \scriptsize{1006} &	\scriptsize{8} &	\scriptsize{9304154} &	\scriptsize{Duality Symmetric Actions} &	 \scriptsize{229} &	\scriptsize{11} \\
\hline
 \end{tabular}
 }
\caption{List of articles with highest total DC scores for $\alpha=0$ and $\alpha=0.2$ along with their number of citations (\#C) and PageRank (PR) rank.}
\label{tbl:total_influence}
\end{table*}
}
Table~\ref{tbl:total_influence} lists ten articles with  highest DC scores over the entire time period along with these articles total citations count and rank according to PageRank, also computed over the entire time period. The top-10 list at $\alpha=0.0$ is relatively insensitive to the value of $\gamma$, with only two articles 9908142 and 9906064 moving out of the top-10 position as $\gamma \rightarrow 1.0$. For this value of $\alpha$, DC takes number of citations into account only, and indeed the list contains articles with the highest citations counts, which are reported in column \#C.

In addition to direct citations, DC allows us to take longer citations chains into account. Increasing $\alpha$ to $0.2$ (which corresponds to average citations chain of length 1.25) dramatically alters the rankings. Recent papers drop in rankings since not enough time had passed to create longer citations chains to them. For example, article 9711200 that was ranked 1 moves to position 103. Other papers with far fewer citations, $\sim 100$, move to the top of the list. As $\gamma$ increases to it optimal value, three papers 9410167, 9510017, and 9510135 are replaced in the top-10 list by three new papers (9209016, 9208055, 9303057). Remarkably, two of them are by the same author, A. Sen.

In summary dynamic centrality leads to a completely different view of importance than citations count and PageRank. Only nine of the 20 articles rated highest by PageRank appear among the top-20 articles rated highest by DC (using optimal parameter values). Another striking difference is that Edward Witten authored five of the 20 articles ranked highest by PageRank, while Ashoke Sen authored four. Among the 20 articles rated highest by DC, Ashoke Sen appears as an author seven times and Ed Whitten two times. While Sen may not be as famous as Whitten, he is a major figure in string theory, who had a remarkable ability to write prescient papers~\cite{Strassler}. He is also a prolific author, fifth most productive one in the \emph{arXiv} data set. Dynamic centrality is able to discover ``hidden gems'' by this influential physicist which are overlooked by other metrics.

\section{Conclusion}
\label{sec:conclusion}
We have presented a novel formulation of centrality for dynamic networks that measures the number of paths that exist over time in a network.
Given snapshots of the network at different times showing the connected nodes, we can calculate dynamic centrality and use this metric to rank nodes by how well connected they are over time to the rest of the network. In addition, we can identify nodes that are best connected to, and therefore, exert most influence on, a given node. We can also vary the time and length scale parameters to identify nodes that are globally or locally connected.

Dynamic centrality gives a different view of importance in a network than other measures, such as static centrality and PageRank. We illustrated the differences on an example network. In addition, we applied dynamic centrality to study scientific papers citations network. Even though this data set has been extensively studied in the past, we were able to discover interesting new facts, including an influential articles that were overlooked by other approaches.

Citations networks are limited in their dynamics, since edges can only appear, but never disappear. We plan to apply our approach to more general dynamic networks.

\small
\bibliographystyle{abbrv}

\end{document}